# Generation of atomic entangled states using linear optics

Zhuo-Liang Cao[*], Ming Yang

Department of Physics, Anhui University, Hefei, 230039, PRChina

## Abstract

In this paper, we propose a novel scheme that can generate two-atom maximally entangled states from pure product states and mixed states using linear optics. Because the scheme can generate pure maximally entangled states from mixed states, we denote it as purification-like generation scheme.

---
[*] E-mail address: caoju@mars.ahu.edu.cn

Quantum superposition principle is a fundamental principle in quantum physics. When used in composite system, it can induce an entirely new result, which is different from the classical physics, i.e. quantum entanglement [1]. After a long time debate on the completeness of quantum mechanics between Niels Bohr and Albert Einstein, the existence of entanglement between two systems, whose state can not be expressed as the product state of the two systems, has been proved [1]. In this sense, the quantum entanglement is used to disprove the local hidden variable theory [2]. Because of the non-locality feature of entanglement, entangled states have been widely used in quantum information processing, such as quantum cryptography [3], quantum computer [4], and quantum teleportation [5].

All the above applications are based on the entangled states, so the generation of entangled states plays a critical role in quantum information processing. Many theoretical and experimental schemes for the generation of entangled states have been proposed in Cavity QED [6], ion trap [7], and NMR [8].

In photonic case, the polarization entangled photons have been generated in experiment by using Spontaneous Parametric-Down conversion [9]. For atomic case, the schemes for the generation of entangled atomic states have been proposed [6]. Shi Biao-Zheng and Guang-Can Guo have presented a realizable scheme for the generation of entangled atomic states, which is mainly based on the dispersive interaction between atoms and cavity modes. The obvious advantage of it is that the cavity is only virtually excited during the process and the requirement on the cavity quality is greatly loosened, which opens a promising perspective for quantum information processing [10]. This scheme was realized in experiment by the S. Haroche group [11]. Alternatively, the scheme for generation of entangled atomic states using cavity decay has been proposed by M. B. Plenio[12], where the cavity decay plays a constructive role in quantum information processing.

From an experimental point of view, the Cavity QED, ion traps and NMR schemes are all too complicated in experiment when compared with the linear optical ones. Quantum information processing using linear optical elements have been proposed recently [13-14]. Motivated by the simplicity, more and more interests are focused on quantum information using linear optics, so the schemes for generations of atomic states using linear optics have also been proposed [15-16]. Cabrillo, C et al proposed the scheme for generation of entangled states of distant atoms by interference [15]. Lu-Ming Duan et al proposed a novel scheme to entangle two distant atomic ensembles by using the interference of photons, which are emitted from the distant atom ensembles induced by Raman interaction [16].

To simplify the generation process, we should try our best to find the simplest schemes. Motivated by Xing-Xiang Zhou's proposal on non-distortion quantum interrogation [17], we propose a novel and realizable scheme which can generate two-atom maximally entangled states using linear optics. Generally speaking, two atoms are usually in product states or mixed states (evolve from entangled states). In this paper, we will consider the generation process starting from product states and mixed states. For the mixed states case, it looks like an entanglement purification process [14, 18], so we denotes the mixed states case as "purification-like" in this paper.

Here, we will consider two identical atoms, and they are all multi-level system. The level configuration of the atoms has been depicted in Fig.1.

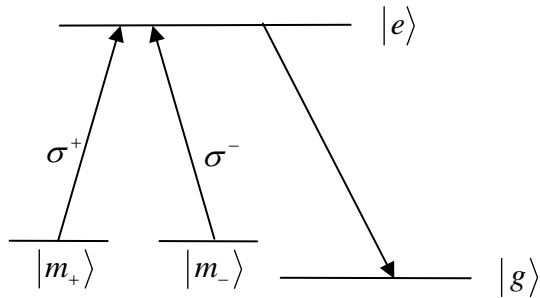

**Fig.1.** The level configuration of the atoms. The atoms in $|m_+\rangle$ (or $|m_-\rangle$) can be excited into the excited state $|e\rangle$ by absorbing one $\sigma^+$ (or $\sigma^-$) polarization photon, and then it will decay to the stable ground state $|g\rangle$ and scatters a photon rapidly. Here, we assume that the decay process is so rapid that the probability of excited emission can be neglected.

Where $|m_+\rangle$ and $|m_-\rangle$ are two degenerate metastable states which are used to store quantum information. $|e\rangle$ is a excited state of atoms and $|g\rangle$ is the stable ground state. Atoms in states $|m_+\rangle$ or $|m_-\rangle$ can be excited into the $|e\rangle$ state by absorbing one + or − circular photons respectively, then it will decay to ground state $|g\rangle$ rapidly and scatter a photon. This process can be expressed as:

$$\hat{a}_\pm^+|0\rangle|m_\pm\rangle \to |S\rangle|g\rangle, \tag{1}$$

where $|S\rangle$ denotes the scattered photons which we assume will not be reabsorbed by the atoms and can be filtered away from the detectors.

The setup for generation of maximally entangled atomic states is depicted in Fig.2.

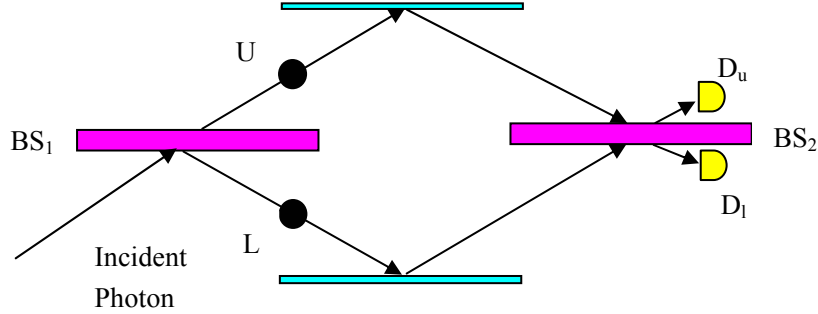

**Fig.2.** The setup for the generation of two-atom entangled states. It is a mach-zehnder interferometer. $BS_1$ and $BS_2$ denote the two beam splitters. U and L denote the two atoms on the upper and the lower port of the interferometer. $D_u$ and $D_l$ are two polarization sensitive single photon detectors at the output upper and lower port.

One Mach-Zehnder interferometer with two beam splitters is the main part of the generation setup. One $\sigma^+$ polarization photon enters the Mach-Zehnder interferometer from the left lower port. If there is no atom on the two arms, the detector at the right upper port will fire, while the right lower port detector will not file. In the case of the existence of one atom in arbitrary superposition states of $|m_+\rangle$ and $|m_-\rangle$ at each arm of the Mach-Zehnder interferometer, the upper and the lower port detectors all have the probability of fire. If we select the superposition coefficients of the initial states of the two atoms appropriately, we can get the maximally entangled atomic states conditioned on the fire at $D_l$.

Next, we will analyze the process in details. We suppose that the two atoms ($U$, $L$) are initially prepared in the following states:

$$|\Psi\rangle_U = \alpha|m_+\rangle_U + \beta|m_-\rangle_U, \quad (2a)$$

$$|\Psi\rangle_L = a|m_+\rangle_L + b|m_-\rangle_L. \quad (2b)$$

where the coefficients $\alpha, \beta, a, b$ satisfy $|\alpha|^2 + |\beta|^2 = 1$ and $|a|^2 + |b|^2 = 1$. These states can be prepared by a laser pulse, and the coefficients can be modulated by the amplitude of the laser pulse.

The effect of the beam splitter on the input photon can be expressed as:

$$\hat{a}^+_{l,\pm}|0\rangle \xrightarrow{BS} \frac{1}{\sqrt{2}}(\hat{a}^+_{u,\pm} + i\hat{a}^+_{l,\pm})|0\rangle, \quad (3a)$$

$$\hat{a}^+_{u,\pm}|0\rangle \xrightarrow{BS} \frac{1}{\sqrt{2}}(\hat{a}^+_{l,\pm} + i\hat{a}^+_{u,\pm})|0\rangle. \quad (3b)$$

That is to say, the beam splitter takes no effect on the polarization of the input photon, and reflects the wave function with a $\frac{\pi}{2}$ phase shift.

The two atoms can be placed on the two arms of the Mach-Zehnder interferometer by using the trapping techniques [19].

Next, we will trace the input photon and give the evolution of the total system. After one $\sigma^+$ polarization photon entering the left lower port of the Mach-Zehnder interferometer, its wave function will be split into two parts (the upper arm and the lower arm) by the first beam splitter($BS_1$). Because the two atoms are placed on the two arms, they will interact with the different parts of the wave function. Then the two parts of the wave function will be combined by the second beam splitter ($BS_2$). The total evolution of the system can be expressed as follow:

$$\hat{a}^+_{l,+}|0\rangle(\alpha|m_+\rangle_U + \beta|m_-\rangle_U)(a|m_+\rangle_L + b|m_-\rangle_L)$$

$$\rightarrow \frac{1}{\sqrt{2}}\alpha|S\rangle_U|g\rangle_U(a|m_+\rangle_L + b|m_-\rangle_L) + \frac{i}{\sqrt{2}}a|S\rangle_L|g\rangle_L(\alpha|m_+\rangle_U + \beta|m_-\rangle_U)$$

$$+ \frac{i}{2}\hat{a}^+_{u,+}|0\rangle(\beta a|m_-\rangle_U|m_+\rangle_L + \alpha b|m_+\rangle_U|m_-\rangle_L + 2\beta b|m_-\rangle_U|m_-\rangle_L)$$

$$+\frac{1}{2}\hat{a}_{l,+}^{+}|0\rangle\bigl(\beta a|m_{-}\rangle_{U}|m_{+}\rangle_{L}-\alpha b|m_{+}\rangle_{U}|m_{-}\rangle_{L}\bigr). \tag{4}$$

From the above result, we can get that the two atoms will be left in three possible states corresponding to three measurement results on the two output ports respectively, which is depicted in the following table:

| measurement results | system states | Probability of obtaining it |
|---|---|---|
| $D_l$ fires | $\frac{1}{2}\hat{a}_{l,+}^{+}|0\rangle\begin{pmatrix}\beta a|m_{-}\rangle_{U}|m_{+}\rangle_{L}\\-\alpha b|m_{+}\rangle_{U}|m_{-}\rangle_{L}\end{pmatrix}$ | $\frac{1}{4}\bigl(|\beta a|^2+|\alpha b|^2\bigr)$ |
| $D_u$ fires | $\frac{i}{2}\hat{a}_{u,+}^{+}|0\rangle\begin{pmatrix}\beta a|m_{-}\rangle_{U}|m_{+}\rangle_{L}\\+\alpha b|m_{+}\rangle_{U}|m_{-}\rangle_{L}\\+2\beta b|m_{-}\rangle_{U}|m_{-}\rangle_{L}\end{pmatrix}$ | $\frac{1}{4}\bigl(|\beta a|^2+|\alpha b|^2+4|\beta b|^2\bigr)$ |
| No fire at $D_l$ and $D_u$ | $\frac{1}{\sqrt{2}}\alpha|S\rangle_{U}|g\rangle_{U}\bigl(a|m_{+}\rangle_{L}+b|m_{-}\rangle_{L}\bigr)$ $+\frac{i}{\sqrt{2}}a|S\rangle_{L}|g\rangle_{L}\bigl(\alpha|m_{+}\rangle_{U}+\beta|m_{-}\rangle_{U}\bigr)$ | $\frac{1}{2}\bigl(|\alpha|^2+|a|^2\bigr)$ |

**Table**1. The atoms will be left in three possible states with its own possibility corresponding to the measurement result.

After evolution, single photon polarization measurements will be operated at the two output ports. The table tells us that if the $D_l$ fires, we get two-atom entangled states:

$$|\Psi\rangle_{UL}=\frac{1}{2}\sqrt{|\beta a|^2+|\alpha b|^2}\left(\frac{\beta a}{\sqrt{|\beta a|^2+|\alpha b|^2}}|m_{-}\rangle_{U}|m_{+}\rangle_{L}-\frac{\alpha b}{\sqrt{|\beta a|^2+|\alpha b|^2}}|m_{+}\rangle_{U}|m_{-}\rangle_{L}\right). \tag{5}$$

If we modulate the coefficients of the initial states to make $\alpha,\beta,a,b$ satisfy $|\alpha|=|a|$ and $|\beta|=|b|$, the two atoms can be left in maximally entangled state $|\Psi\rangle_{UL}=\frac{1}{\sqrt{2}}\bigl(|m_{-}\rangle_{U}|m_{+}\rangle_{L}-|m_{+}\rangle_{U}|m_{-}\rangle_{L}\bigr)$ with probability $P=\frac{1}{2}|a|^2\bigl(1-|a|^2\bigr)$. From this analysis, we conclude that the two atoms must be prepared in the same superposition state or there is only a phase difference between them, then we can get the two-atom

maximally entangled state. The successful probability is a function of the modulus of the initial states.

If the $D_u$ fires, the two atoms will be left in the following state:

$$|\Psi\rangle_{UL} = \beta|m_-\rangle_U (a|m_+\rangle_L + b|m_-\rangle_L) + b(\alpha|m_+\rangle_U + \beta|m_-\rangle_U)|m_-\rangle_L. \quad (6a)$$

Conditioned on the first case condition, it will reduce to:

$$|\Psi\rangle_{UL} = b|m_-\rangle_U (a|m_+\rangle_L + b|m_-\rangle_L) + b(a|m_+\rangle_U + b|m_-\rangle_U)|m_-\rangle_L. \quad (6b)$$

It is a partially entangled state, and can be converted into the standard entangled state.

The last result (no detector fires) gives no contribution to the entanglement of the atomic state.

The above analyses are all based on the fact that the two atoms are placed on the two arms of the Mach-Zehnder interferometer. Because the present Mach-Zehnder interferometer is still a local apparatus, we can only generate the entangled atomic states locally. However, we can extend the length of the arms of the interferometer, and the polarized photon can propagates in a polarization-preserving fiber over a long distance without destruction on the polarization of the photon. So we can prepare the maximally entangled states between remote locations, which will become a robust resource for quantum communication [5, 20]. Most of the previous preparation schemes prepare the entangled states at one location, and then the entangled particles will be distributed among different users for quantum communication purpose. But during the transmission of the particles, it will unavoidably couple with environments, and then the entanglement will degrade exponentially. So the entangled states after distribution is usually mixed ones, which need the purification process [14, 18, and 21] before use. Here, we use photons as flying bit, which avoids the above difficulty. So the fidelity of the generated states has been enhanced to a good level.

In this scheme, the probability of obtaining the maximally entangled state is relatively small. If we start the generation scheme with two atoms in same product state, the maximal successful probability is only $\frac{1}{8}$. At the same time, the two atoms at two remote locations are probably in a mixed state, because the two atoms can be from another entangled resource, and the entanglement of them degraded from the environment during transmission. So we will consider the generation of entangled states from the mixed states. Then we find that the successful probability of the mixed states case can be larger than the product states case. Furthermore, the generated maximally entangled states are pure ones, that is to say, we get pure maximally entangled states from mixed states. In this sense, the scheme for mixed states looks like an entanglement purification one [18].

Suppose that the initial mixed state is in the following form [14]:

$$\rho_{UL} = F|\Psi^+\rangle_{UL}\langle\Psi^+| + (1-F)|\Phi^+\rangle_{UL}\langle\Phi^+|, \tag{7}$$

where $|\Psi^+\rangle_{UL} = \frac{1}{\sqrt{2}}(|m_+\rangle_U|m_-\rangle_L + |m_-\rangle_U|m_+\rangle_L)$, $|\Phi^+\rangle_{UL} = \frac{1}{\sqrt{2}}(|m_+\rangle_U|m_+\rangle_L + |m_-\rangle_U|m_-\rangle_L)$ are two Bell states.

To express the evolution clearly, we will consider the mixed state as the probabilistic mixture of pure two-atom entangled states, i.e. the state $|\Psi^+\rangle_{UL}$ with probability $F$ and the state $|\Phi^+\rangle_{UL}$ with probability $1-F$.

For the $|\Psi^+\rangle_{UL}$ case, the evolution can be expressed as:

$$\hat{a}_{l,+}^+|0\rangle\frac{1}{\sqrt{2}}(|m_+\rangle_U|m_-\rangle_L + |m_-\rangle_U|m_+\rangle_L)$$

$$\rightarrow \frac{1}{2}(|S\rangle_U|g\rangle_U|m_-\rangle_L + i|m_-\rangle_U|S\rangle_L|g\rangle_L) + \frac{i}{2\sqrt{2}}\hat{a}_{u,+}^+|0\rangle(|m_+\rangle_U|m_-\rangle_L + |m_-\rangle_U|m_+\rangle_L)$$

$$+ \frac{1}{2\sqrt{2}}\hat{a}_{l,+}^+|0\rangle(|m_-\rangle_U|m_+\rangle_L - |m_+\rangle_U|m_-\rangle_L). \tag{8a}$$

For the $|\Phi^+\rangle_{UL}$ case, the evolution takes a new form:

$$\hat{a}_{l,+}^+|0\rangle\frac{1}{\sqrt{2}}(|m_+\rangle_U|m_+\rangle_L + |m_-\rangle_U|m_-\rangle_L)$$

$$\rightarrow \frac{1}{2}(|S\rangle_U|g\rangle_U|m_+\rangle_L + i|m_+\rangle_U|S\rangle_L|g\rangle_L)$$

$$+ \frac{i}{\sqrt{2}}\hat{a}_{u,+}^+|0\rangle|m_-\rangle_U|m_-\rangle_L. \tag{8b}$$

If we detect a photon at $D_u$, the two atoms will be left in a mixed state:

$$\rho'_{UL} = \frac{2-F}{4}\left(\frac{2(1-F)}{2-F}|m_-\rangle_U|m_-\rangle_L\langle m_-|_U\langle m_-|_L + \frac{F}{2-F}|\Psi^+\rangle_{UL}\langle\Psi^+|\right), \qquad (9)$$

whose fidelity (for $|\Psi^+\rangle_{UL}$) is lower than the initial one. So we consider this result as garbage. If we detect one photon at $D_l$, the two atoms are left in a pure maximally entangled state $|\Psi\rangle_{UL} = \frac{1}{\sqrt{2}}(|m_-\rangle_U|m_+\rangle_L - |m_+\rangle_U|m_-\rangle_L)$ with probability $P' = \frac{F}{4}$. So long as the initial fidelity satisfies $F > \frac{1}{2}$, the successful probability of the mixed states case will be larger than the pure product states case. This point can be understood easily. The pure states case is starting from a product state, but the mixed states case from a partially entangled state. Naturally, the probability of later case is larger than the former one.

This case (mixed states) looks like an entanglement purification process, because we have lifted the fidelity of the initial state. But, in a strict definition [18], it is still a generation process. The entanglement purification process involves only local operations and classical communication. Here, the operations we use are joint ones. But, this scheme can transform mixed states into pure maximally entangled ones, which can not be realized by the previous generation and purification schemes. It is a novel advantage.

The present scheme requires that the two atoms are accurately placed on the two arms of the Mach-Zehnder interferometer. Although the trapping of one atom is possible, there is always an error from the experimental point of view. What will happen if the two atoms are not exactly placed on two arms? There are three possible errors. That is to say, there is an error on each arm, on the upper detector or on the lower detector. Because the result of the no fire case gives no contribution to the entangled states, the first error takes no effect on the final result. Through analysis, if the latter two errors occur, we probably can not get the entangled states of the two atoms. So the main purpose of this scheme is to place the atoms on the two arms of the Mach-Zehnder interferometer accurately.

In conclusion, we propose a scheme that can entangle the two locally isolated or remote isolated atoms and "purify" mixed states using linear optics probabilistically. The obvious advantage of our scheme is that it is simpler than the previous generation and purification schemes. In addition, the scheme can "purify" maximally entangled states from mixed states, which is novel and can not be realized by the previous generation and purification processes.

## Acknowledgements

We thank Zheng-Wei Zhou for his useful discussion. This work is supported by Anhui Provincial Natural Science Foundation under Grant No: 03042401 and the Key Program of the Education Department of Anhui Province.